\documentclass[journal=jctcce,manuscript=article,layout=traditional]{achemso}

\usepackage{graphicx}
\usepackage{caption}
\usepackage{subcaption}
\usepackage{bm}
\usepackage[colorlinks=true,breaklinks=true,linkcolor=red,citecolor=magenta,urlcolor=blue]{hyperref}
\usepackage{comment}
\usepackage{amsmath}
\usepackage{amsfonts,amssymb}
\usepackage{booktabs}
\usepackage{multirow}
\usepackage{multicol}

\usepackage[utf8]{inputenc}
\usepackage[T1]{fontenc}
\usepackage{qcircuit}
\usepackage{float}

\usepackage[version=4]{mhchem}
\usepackage{braket}

\title{A scalable route to first-order response properties with correlated sampling phaseless auxiliary-field quantum Monte Carlo}

\author{Leon Otis}
\affiliation{Department of Chemistry, Rice University, Houston, TX 77005-1892, USA}
\author{Sri Gudivada}
\affiliation{Department of Chemistry, Rice University, Houston, TX 77005-1892, USA}
\author{Marvin Friede}
\affiliation{Mulliken Center for Theoretical Chemistry, University of Bonn, 53115 Bonn, Germany}
\author{James Shee}
\affiliation{Department of Chemistry, Rice University, Houston, TX 77005-1892, USA}
\alsoaffiliation{Department of Physics and Astronomy, Rice University, Houston, TX 77005-1892, USA}
\email{james.shee@rice.edu}

\begin{document}

\begin{tocentry}

\includegraphics[width=8.25cm,height=4.45cm]{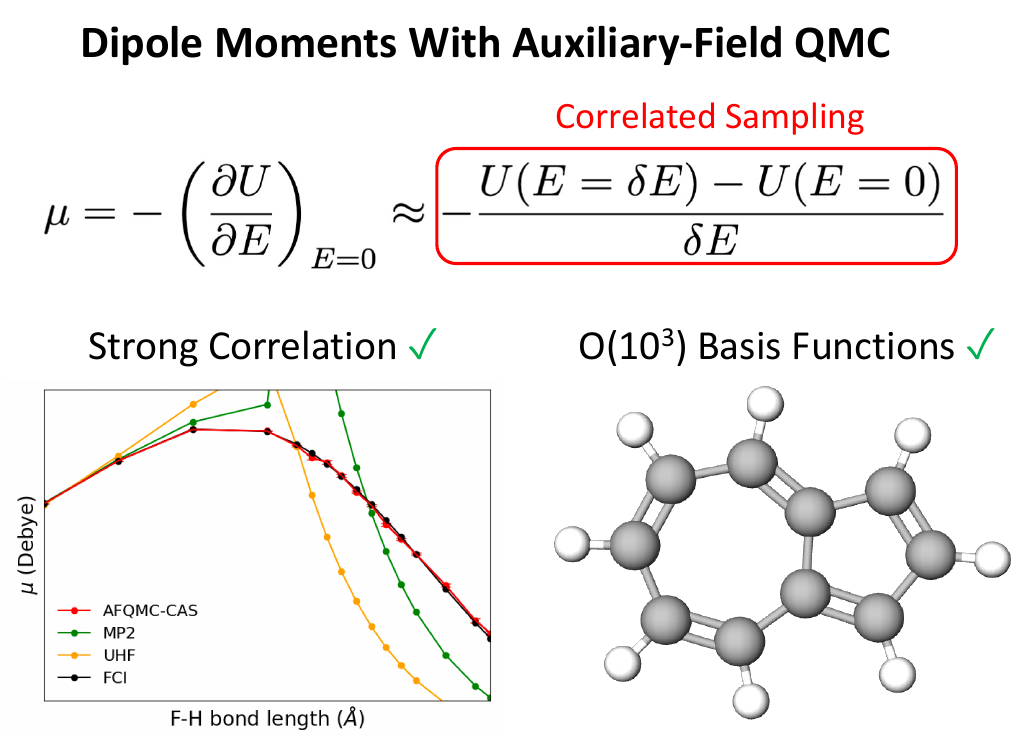}

\end{tocentry}

\begin{abstract}
To make useful connections with experimental measurements, correlated electronic structure theories must accurately predict chemical properties in addition to energies.  We present a finite-difference based algorithm to compute first-order response properties with phaseless auxiliary-field quantum Monte Carlo (ph-AFQMC) that relies on a branching correlated sampling approach.  Focusing on electric dipole moments, we show that mean-field trial wave functions are sufficient to obtain high accuracy relative to CCSD(T) and experimental measurements for a set of 21 molecules, ranging in size from 2 to 18 atoms in their equilibrium geometries.  As with energies, the quality of predicted dipole moments can be systematically improved with the use of correlated trial wave functions, even (or especially) in strongly correlated regimes.  We show that the challenges faced by low-order perturbation theories in predicting the dipole moment of hydrogen fluoride across its dissociation coordinate are overcome with ph-AFQMC when using relatively simple trials.  The key advantage of our approach over those previously reported for ph-AFQMC is its scalability to large system sizes with a phaseless bias no worse than that of a typical ground-state energy calculation; we routinely converge dipole moments for systems with more than one thousand basis functions. 
\end{abstract}

\maketitle







The accurate calculation of physical properties, typically related to the response of a molecule or material to weak electromagnetic fields, remains a significant challenge for approximate electronic structure methods.
In the context of semi-empirical density functional theory development, substantial effort has been devoted to obtaining accurate energies, at times at the expense of accuracy for other observables, e.g., based on the electron density.\cite{Medvedev2017}
With regard to \emph{ab initio} wavefunction methods, many groups focusing on computing properties other than the energy have developed formalisms such as coupled perturbed Hartree-Fock\cite{Mcweeny1960,Mcweeny1962,Diercksen1966,Wolinski1990} and the lambda equations for coupled cluster gradients and density matrices.\cite{Bartlett2007,Scheiner1987,Salter1989,Gauss1991,Gauss1995} 

Quantum Monte Carlo (QMC) methods, which have 
earned a reputation for highly accurate energetics,\cite{Foulkes2001,Austin2012,Dubecky2016,motta2018ab} face a number of challenges in pursuit of other properties.\cite{Assaraf2000,Assaraf2003,Lopez2019,Tiihonen2021,Hood1997,Kaiser2025}
One type of QMC method, called auxiliary-field QMC (AFQMC) is a promising approach for a variety of correlated many-particle systems.\cite{Zhang2003,motta2018ab,lee2022twenty,Shee2023}
AFQMC calculations of properties, to date, include studies of forces\cite{Motta2018forces,Chen2023_forces}, 
dipole moments\cite{Motta2017,Mahajan2022,Mahajan2023,Mahajan2025}, densities\cite{Chen2021} and correlation functions.\cite{Zhang1997,Motta2014,Vitali2016,Qin2017,Lee2021_phonons,Church2021}
One approach to obtaining properties in AFQMC (including observables corresponding to operators that do not commute with the Hamiltonian) is the back-propagation technique,\cite{Zhang1995} 
which introduces a second imaginary-time propagation of the trial state when computing 
expectation values.
A potential limitation on the accuracy of this approach comes from the 
increased bias caused by the phaseless approximation when a second constrained projection is 
performed.\cite{Motta2017,Lee2021_ekt}
More recently, another route to AFQMC properties has been developed based on automatic differentiation (AD-AFQMC).\cite{Mahajan2023}
Initial applications\cite{Mahajan2023,Mahajan2025} of AD-AFQMC to dipole moments 
have obtained improved agreement with experiment compared to back-propagation, but 
AD-AFQMC is currently limited to systems of a few hundred basis functions due to 
memory constraints.

To overcome these challenges, we introduce an accurate and scalable 
approach to first-order properties in AFQMC based on finite differences (FD-AFQMC).
This strategy rests on the use of correlated sampling to resolve small energy differences with a tractable number of Monte Carlo samples.  In this study we focus on the computation of electric dipole moments, and demonstrate the accuracy of FD-AFQMC across a range of molecular 
sizes, including systems currently inaccessible to AD-AFQMC and canonical CCSD(T).

We first briefly review the main aspects of AFQMC before discussing the use of 
correlated sampling and finite differences.
AFQMC is a projective method based on the application of the imaginary time 
propagator to some initial state $\ket{\psi_I}$:

\begin{equation}
    \ket{\psi_0} = \lim_{\tau \to\infty} e^{-\tau \hat{H}} \ket{\psi_I}.
\end{equation}
The imaginary time propagation is discretized into small steps, $\Delta\tau$,
and the Trotter-Suzuki decomposition\cite{Suzuki1976} is applied to the 
exponential of the Hamiltonian to separate the one- and two-body parts.  The electron-electron interaction $\hat{V}$ can be written as a sum of squares of one-body operators:
\begin{equation}
    \hat{V} = \frac{1}{2} \sum_{\gamma} \hat{L}_\gamma^2,
\end{equation}
where 
\begin{equation}
    \hat{L}_\gamma = \sum_{pq} L_{pq}^{\gamma} \hat{a}_p^\dagger \hat{a}_q,
\end{equation}
$\hat{a}_p^\dagger$ and $\hat{a}_q$ are the usual one-electron creation and 
annihilation operators with $p$ and $q$ as orbital indices.
The $\hat{L}_\gamma$ can usually be obtained with a modified Cholesky decomposition\cite{Pedersen2024} and 
introducing them enables the exponential of the interaction term to be 
rewritten using the Hubbard-Stratonovich transformation:\cite{Hubbard1959,Stratonovich1957}

\begin{equation}
    e^{-\Delta\tau\hat{V}} = \int \exp[\sqrt{-\Delta\tau}\sum_{\gamma}x_\gamma \hat{L}_\gamma] \prod_\gamma\frac{\exp(-\frac{1}{2}x_\gamma^2)}{\sqrt{2\pi}}dx_\gamma
\end{equation}
The auxiliary fields $x_\gamma$ introduced by the transformation can be 
sampled from a normal distribution, allowing a stochastic realization of the
projection with random walkers.
In practice, the AFQMC projection is often constrained by the phaseless approximation imposed by a precomputed trial wave function.\cite{Zhang2003} 
The phaseless approximation avoids numerical instabilities from the 
phase problem by enforcing that the walker weights remain real-valued and positive.  Any potential 
bias from the constrained projection can be mitigated in principle by systematically improving the quality of the trial.

Correlated sampling procedures seek to reduce statistical noise in the 
differences of stochastically estimated quantities.\cite{Kalos2008}
At a very general level, when computing a Monte Carlo estimate of the difference 
of two quantities $A-B$, the variance of the difference is:

\begin{equation}
    \text{var}(A-B) = \text{var}(A) + \text{var}(B) -2 \text{Cov}(A,B) = \text{var}(A) + \text{var}(B) -2(\langle AB \rangle - \langle A\rangle \langle B\rangle ).
\end{equation}
The covariance $\text{Cov}(A,B)$ is zero if $A$ and $B$ are independently sampled, 
which maximizes $\text{var}(A-B)$ while in the opposite limit of perfect 
correlation in the sampling, $2\text{Cov}(A,B) = \text{var}(A) + \text{var}(B)$ 
and $\text{var}(A-B)$ is zero with no noise in the estimate of $A-B$.

In the context of AFQMC, correlated sampling is implemented by using the same 
set of auxiliary fields and pairing the propagation of 
walkers for two (or more) similar systems.\cite{Shee2017}
Recent work\cite{Chen2023_cs} has shown that population control of the walkers can be approximately incorporated in a correlated fashion through multiple schemes. 
In this work, we use the ``static" approach 
where the branching and annihilation importance sampling decisions based on the walker weights are based on a primary AFQMC calculation and used for the other correlated 
secondary calculations.
The use of correlated sampling in AFQMC remains at a relatively early stage with some initial applications to ionization potentials\cite{Shee2017}, electron binding energies,\cite{Hao2018} diatomic dissociation energies,\cite{shee2019achieving} nuclear forces,\cite{Chen2023_cs,Goings2025} and noncovalent interactions.\cite{Awasthi2025}

The dipole moment can be expressed as a derivative of the energy $U$ with respect 
to electric field $E$:

\begin{equation}
    \mu = -\left(\frac{\partial U}{\partial E}\right)_{E=0} \approx -\frac{U(E=\delta E) - U(E=0)}{\delta E}.
\end{equation}
The derivative can be approximated with a difference of energies between a zero-field calculation and one with small field $\delta E$, and this approach can be 
repeated for each component of the dipole moment in general.
In principle, the finite-difference approximation of the energy derivative improves as the magnitude of the perturbing field $\delta E$ goes to 
zero; however, in this limit the energy difference in the numerator will be small and the
suppression of statistical noise becomes critical. This motivates our present use of correlated sampling; indeed, we find that finite-difference dipole moments cannot be converged without it.


To summarize methodological details of the calculations to follow, we employ an 
electric field strength of $10^{-5}$ a.u. unless stated otherwise.  
With a few exceptions, we use restricted Hartree-Fock (RHF) trial wave functions obtained from PySCF\cite{sun2018pyscf,sun2020recent}, which we also use for HF, MP2, and 
coupled cluster dipole moment calculations.
The frozen-core approximation is used throughout our calculations.
We also compute a set of dipole moments from DFT using ORCA\cite{ORCA,ORCA6} 
to compare against our results from wavefunction methods.
Our geometries are such that the molecular center of charge is at the origin.  In cases where the orientation of the dipole moment is clear from 
symmetry, we only need to use an electric field perturbation in that direction.
When we consider larger and less symmetric molecules, we have perturbations along the x, y, and z 
directions along with the unperturbed case for a total of four AFQMC calculations.
In our results, we show the magnitude of the dipole moment vector, computed from the set of Cartesian components, and it is this quantity that is compared with different levels of theory and experimental measurements.
To characterize the overall deviation of AFQMC values from reference values for a set of molecules we compute the root mean squared regularized error (RMSRE)\cite{Hait2018} defined as:
\begin{equation}
    \text{RMSRE} =\sqrt{\frac{1}{N} \sum_{i=1}^N \big(\frac{\mu_{i} - \mu_{i,\text{ref}}}{\max(\mu_{i,\text{ref}}, 1\,\text{D})}\big)^2}.
\end{equation}

As a first check of the accuracy of our methodology, we consider a set of 
small and highly symmetric molecules that have previously been studied with 
AD-AFQMC.\cite{Mahajan2025}
As shown in Table \ref{tab:small_molecules_data}, we see that our FD-AFQMC technique is quite accurate with dipole moments often within a 
few hundredths of a Debye of the values from automatic differentiation 
as well as experiment.
These results were obtained with modest computational effort, using only 512 walkers propagating for 40 Ha$^{-1}$. 
We also note that we have found it completely infeasible to resolve the required energy differences with uncorrelated sampling even in these small systems, highlighting the importance of correlated sampling to our finite-difference strategy.

\begin{table}[H]

\caption{\footnotesize Dipole moments in Debye for a set of small molecules. Stochastic uncertainties on the last digits are given in parentheses. All calculations use the aug-cc-pVQZ basis set. RHF, AD-AFQMC, and experimental values are taken from Ref. \citenum{Mahajan2025}. }
\begin{tabular}{lllllll}
Method & \ce{H2O} & \ce{NH3} & HCl & HBr & CO & \ce{CH2O} \\ \hline

 RHF & 1.985 & 1.617 & 1.235 & 0.947 & -0.292 & 2.867 \\
AD-AFQMC-RHF & 1.815(5) & 1.479(15) & 1.085(8) & 0.826(8) & 0.025(8) & 2.506(10) \\
FD-AFQMC-RHF & 1.879(5)& 1.491(8) & 1.120(6)  & 0.716(2) &0.033(10) & 2.448(20)\\
Experiment & 1.855\cite{Lide1995crc}& 1.477\cite{Shimizu1970} & 1.093\cite{NIST_diatomic} & 0.826\cite{NIST_diatomic} & 0.122\cite{Meerts1977} & 2.333\cite{Theule2003} \\
\end{tabular}

\label{tab:small_molecules_data}
\end{table}




Now we turn 
our attention to computing dipole moments of larger molecules, including 
those beyond the current reach of AD-AFQMC.
To devise a manageable set of test molecules, we have considered 
experimental values listed in the Computational Chemistry Comparison and Benchmark Database (CCCBDB).\cite{NIST_cccbdb}
We selected molecules with more than 12 atoms with listed experimental 
uncertainties of under 0.05 Debye.
These experimental values can also be found in various primary literature sources cited on the CCCBDB page.\cite{NIST_hydrocarbon,Mehrotra1977,Simmons1981,Hüttner1989,Hellwege_1974,Kato1980,NAP_dielectric,Nugent1962,Kitchin1975,Parkin1981,Caminati1993,Huber2005}
We performed DFT geometry optimizations of our chosen molecules using the $\omega$B97M-V functional\cite{Mardirossian2016} and the def2-QZVPP basis set with ORCA.\cite{ORCA,ORCA6}
For FD-AFQMC calculations on this set, 2048 walkers were propagated for 40 Ha$^{-1}$ in all cases.

\begin{figure}[H]
    \centering
    \includegraphics[width=\linewidth]{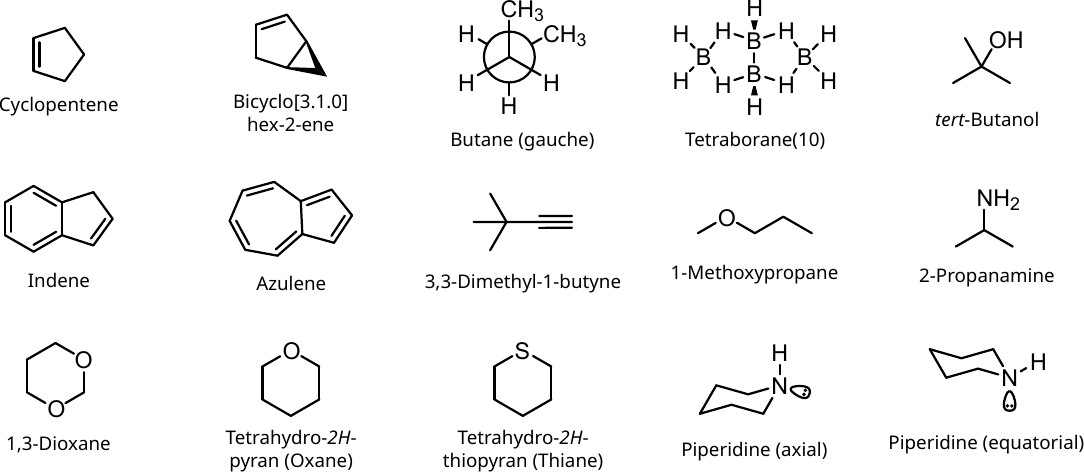}
    \caption{\footnotesize Test set of larger molecules from CCCBDB\cite{NIST_cccbdb} with experimentally measured dipole moments.}
    \label{fig:structures}
\end{figure}

For this set of larger molecules (Figure \ref{fig:structures}), we first compare dipole moments computed from different theoretical methods.
Table \ref{tab:ccpvtz_large_molecules_data} shows good agreement between 
FD-AFQMC and CCSD(T) with differences of only a few hundredths of a Debye except in the case of azulene.
Taking CCSD(T) as the reference, Table \ref{tab:statistics_large_molecules_cc} 
shows that the accuracy of FD-AFQMC across the whole set is very similar to that of CCSD with a RMSRE of about 4 percent,
and can be achieved with $\mathcal{O}(N^4)$ (per sample) scaling with system size vs the $\mathcal{O}(N^6)$ scaling of CCSD.
MP2 performs very well for this test set with an error of only 3 percent in its worst case, but has been found 
to yield significantly larger errors in Ref. \citenum{Hait2018}, especially in the case of spin-polarized systems.  
In terms of absolute errors, FD-AFQMC, CCSD and MP2 are all highly accurate 
with average deviations of only about 0.01 or 0.02 Debye.

\begin{table}[H]
    \caption{\footnotesize Dipole moments in Debye for a collection of larger molecules 
    at different levels of theory. All calculations used a cc-pVTZ basis set. 
    Stochastic uncertainties of FD-AFQMC are given in parentheses.}
    \label{tab:ccpvtz_large_molecules_data}
    \begin{tabular}{lccccc}
        \toprule
        Molecule & FD-AFQMC-RHF  & HF & MP2 & CCSD & CCSD(T) \\
        \midrule
        Cyclopentene & 0.165(4)  & 0.185& 0.178& 0.169& 0.169\\
        2-Propanamine & 1.237(3)  & 1.284& 1.229& 1.232& 1.219\\
        Tetraborane & 0.481(4)  & 0.529 & 0.473& 0.490& 0.478\\
        Gauche Butane & 0.086(3)  & 0.088& 0.088 & 0.088& 0.088\\
        Bicyclo[3.1.0]hex-2-ene & 0.259(6) & 0.249& 0.253& 0.237& 0.242\\
        1,3-Dioxane & 1.998(4)  & 2.229 & 1.986& 2.032& 1.971\\
        1-Methoxypropane & 1.086(4)  & 1.227& 1.073 &1.106 & 1.066\\ 
        \textit{tert}-Butanol & 1.477(4) & 1.628 & 1.473& 1.501& 1.465\\ 
        3,3-Dimethyl-1-butyne & 0.647(6)  & 0.691& 0.568& 0.588& 0.585\\
        Tetrahydro-\textit{2H}-pyran & 1.411(5) & 1.568& 1.402& 1.433& 1.392\\
        Tetrahydro-\textit{2H}-thiopyran & 1.762(7) & 1.933& 1.758& 1.742& 1.722\\
        Equatorial Piperidine &0.822(6) & 0.891 & 0.826 & 0.843& 0.823\\
        Axial Piperidine & 1.110(3)  & 1.193 & 1.115 & 1.129& 1.105\\
        Indene & 0.671(10)  & 0.685& 0.640 & 0.624& 0.620\\
        Azulene & 1.131(30)  & 1.452 & 0.950 & 1.138& 0.981\\
        \bottomrule
    \end{tabular}
\end{table}

\begin{table}[H]
    \caption{\footnotesize Statistics over the set of larger molecules comparing to CCSD(T), including root mean square regularized error, maximum percentage error, mean absolute deviation, and maximum absolute deviation. Absolute deviations are in Debye. All methods use the cc-pVTZ basis set and FD-AFQMC uses a RHF trial.} 
    \label{tab:statistics_large_molecules_cc}
    \begin{tabular}{lcccc}
        \toprule
        Method & RMSRE (\%) & MaxE  (\%) &MeanAD (D) & MaxAD (D) \\
        \midrule
        FD-AFQMC-RHF   & 4.14 & 13.22 & 0.028& 0.132\\
        RHF            & 14.99 & 47.12 & 0.127 & 0.471\\
        MP2          & 1.33 & 3.08  & 0.013 &0.036\\
        CCSD           & 4.451 & 15.68 & 0.029 & 0.157\\
        \bottomrule
    \end{tabular}
\end{table}

\begin{figure}[H]
    \centering
    \includegraphics[width=\linewidth]{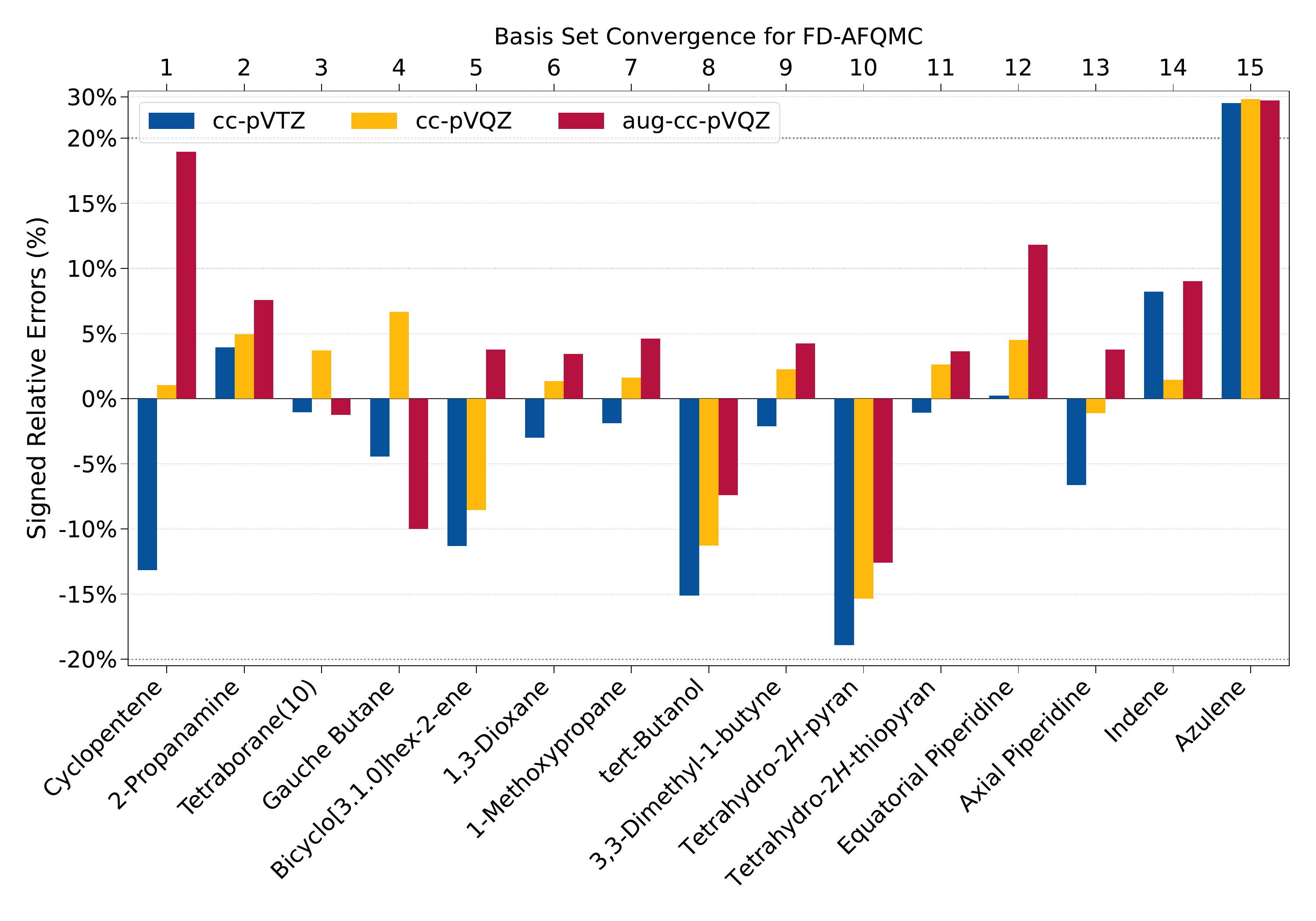}
    \caption{\footnotesize Signed relative error of FD-AFQMC-RHF to experiment in different basis sets.}
    \label{fig:basis_afqmc}
\end{figure}

We also compare FD-AFQMC against experiment and consider the effects of basis set 
size.
As seen in Figure \ref{fig:basis_afqmc}, while the use of larger basis sets can 
increase the dipole moment predicted by FD-AFQMC, there is not a consistent 
trend across all the molecules in our test set.
Some cases with larger errors at the cc-pVTZ level, such as tert-butanol and tetrahydro-2H-pyran, are noticeably improved by increasing the basis set to aug-cc-pVQZ.
However, use of aug-cc-pVQZ can lead to overestimation of the dipole moment and 
worsen the level of agreement with experiment, most prominently in the cases of 
cyclopentene and equatorial piperidine.
Besides accuracy considerations, our results are also a demonstration of our 
method's ability to scale to large numbers of basis functions. 
Our largest molecule, azulene, has 412 basis functions with cc-pVTZ and 1168 
with aug-cc-pVQZ, well beyond the current capabilities of the automatic differentiation approach.

\begin{figure}[H]
    \centering
    \includegraphics[width=0.7\linewidth]{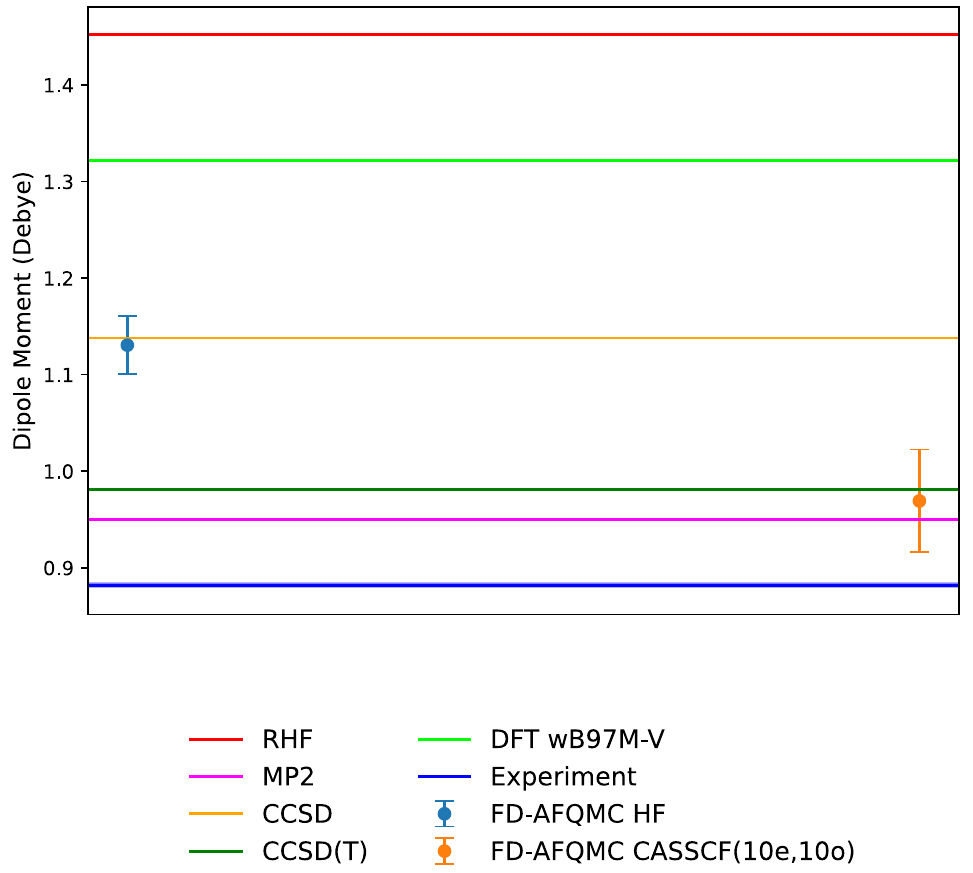}
    \caption{\footnotesize Dipole moment of azulene. The DFT calculation used a def2-QZVPP basis set  while all others used cc-pVTZ. $99\,\%$ weight of the CAS wave function corresponding to 726 determinants was used in the AFQMC trial.}
    \label{fig:azulene_cas}
\end{figure}

In both absolute and percentage terms, the errors of FD-AFQMC relative to experiment for our large molecule set are encouragingly small. 
In Table \ref{tab:aug_vqz_exp_large_molecules}, aug-cc-pVQZ values typically differ 
from experiment by less than 0.1 D and the largest error is 0.256 D in the 
case of azulene.
The percentage error across the entire set in Table \ref{tab:statistics_exp} for FD-AFQMC is comparable to those of the most 
successful DFT functionals we have tested (note the high accuracy of the hybrid functionals B3LYP and PBE0), and somewhat better than those of 
$\omega$B97X-V and $\omega$B97M-V.
For azulene, where the choice of basis set has little effect, we investigate the role of the trial function in AFQMC and apply 
our finite-difference approach with a CASSCF wave function from a (10e,10o) active space of $\pi$ and $\pi^*$ orbitals.
We see in Figure \ref{fig:azulene_cas} that a better trial wave function leads to a 
significant improvement in the dipole moment to a value of 0.969(5) D, within 0.1 D of the experimental value.
The azulene results with RHF and CASSCF trials provide a demonstration of the 
systematic improvability of our approach to dipole moments. 

\begin{table}[H]
    \caption{\footnotesize Dipole moments in Debye for the collection of larger molecules. FD-AFQMC values use the aug-cc-pVQZ basis set and RHF trial. Stochastic and experimental uncertainties are in parentheses.}
    \label{tab:aug_vqz_exp_large_molecules}
    \begin{tabular}{lll}
        \toprule
        Molecule & FD-AFQMC & Experiment  \\
        \midrule
        Cyclopentene & 0.226(4)& 0.190(6)\cite{NIST_hydrocarbon}  \\
        2-Propanamine & 1.280(6)& 1.190(30)\cite{Mehrotra1977} \\
        Tetraborane & 0.480(7) & 0.486(2)\cite{Simmons1981} \\
        Gauche Butane & 0.081(4) & 0.090(2)\cite{Hüttner1989} \\
        Bicyclo[3.1.0]hex-2-ene & 0.303(6) &0.292(15)\cite{NIST_hydrocarbon} \\
        1,3-Dioxane & 2.131(8) & 2.060(20)\cite{Hellwege_1974}\\
        1-Methoxypropane & 1.158(6)& 1.107(13)\cite{Kato1980} \\ 
        \textit{tert}-Butanol & 1.611(8) & 1.740(40)\cite{NAP_dielectric}\\ 
        3,3-Dimethyl-1-butyne & 0.689(5) & 0.661(4)\cite{Nugent1962}\\
        Tetrahydro-\textit{2H}-pyran & 1.521(4) & 1.740(20)\cite{Hellwege_1974}\\
        Tetrahydro-\textit{2H}-thiopyran & 1.846(4) & 1.781(10)\cite{Kitchin1975}\\
        Equatorial Piperidine & 0.917(6) & 0.820(20)\cite{Parkin1981}\\
        Axial Piperidine & 1.234(8) & 1.189(15)\cite{Parkin1981}\\
        Indene & 0.676(14) & 0.620(20)\cite{Caminati1993}\\
        Azulene & 1.138(33) & 0.882(2)\cite{Huber2005}\\
        \bottomrule
    \end{tabular}
\end{table}

\begin{table}[H]
    \caption{\footnotesize Statistics over the set of larger molecules comparing to experiment, including root mean square regularized error and maximum percentage error. FD-AFQMC values use the aug-cc-pVQZ basis and RHF trial, while DFT calculations use the def2-QZVPP basis.} 
    \label{tab:statistics_exp}
    \begin{tabular}{lcc}
        \toprule
        Method & RMSRE  (\%) & MaxE (\%) \\
        \midrule
        FD-AFQMC-RHF   & 8.70 & 25.60  \\
        PBE            & 8.86 & 20.17  \\
        B3LYP          & 7.02 & 14.58  \\
        PBE0           & 8.24 & 19.08  \\
        $\omega$B97X-V & 13.36 & 46.07  \\
        $\omega$B97M-V & 12.67 & 43.82  \\
        \bottomrule
    \end{tabular}
\end{table}

\begin{figure}[H]
    \centering
    \includegraphics[width=\linewidth]{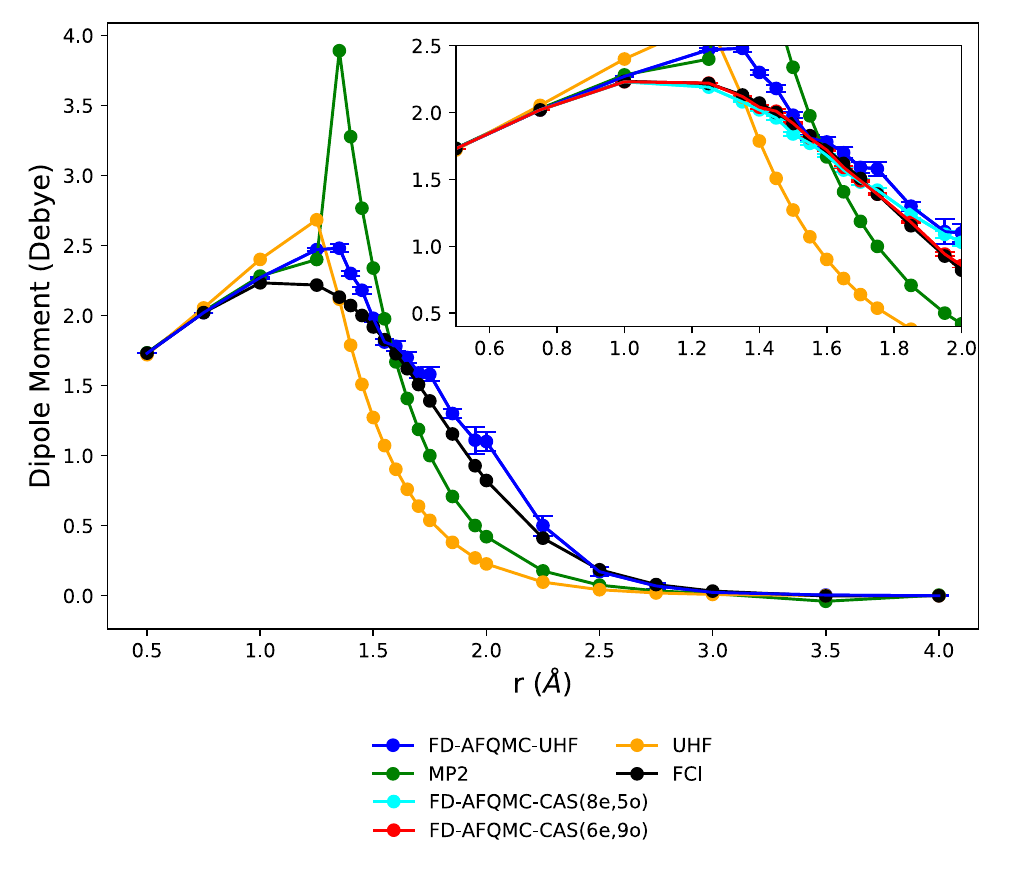}
    \caption{\footnotesize Dipole moments of the hydrogen fluoride diatomic along the dissociation coordinate. All methods use the 6-31G basis set, in order to compare with FCI values. FD-AFQMC-CAS results are shown in the inset. 25 determinants were used for FD-AFQMC-CAS(8e,5o) and 3546 were used for FD-AFQMC-CAS(6e,9o). All FD-AFQMC calculations use 2048 walkers.}
    \label{fig:fh_dipole}
\end{figure}

Given the high accuracy of many methods on our test set, we finally consider FD-AFQMC's performance at non-equilibrium geometries 
with the notoriously challenging dissociation of hydrogen fluoride (FH). 
Previous benchmarking\cite{Hait2018} of DFT functionals on this test system 
has demonstrated that many functionals struggle to accurately predict the dipole moment throughout the dissociation coordinate.
Hybrid functionals such as B3LYP may overestimate the maximum 
dipole moment and predict an overly slow decay at longer separations.
We see in Figure \ref{fig:fh_dipole} that this molecule also 
challenges MP2 with a significant spike in the dipole moment around 1.35 \r{A} followed by an overly rapid decay.
FD-AFQMC with a UHF trial agrees with the FCI reference more closely 
throughout the dissociation than UHF or MP2, but still overestimates 
the dipole moment at intermediate separation lengths.

The limitations of FD-AFQMC-UHF can be alleviated with improved 
trial wave functions, as shown in the inset of Figure \ref{fig:fh_dipole}. 
Trial wave functions of only 25 determinants from (8e,5o) CASSCF 
calculations are sufficient to greatly improve the level of 
agreement between FD-AFQMC and FCI, and any remaining error can be 
eliminated with trials from a larger (6e,9o) active space.
Our ability to converge to the FCI reference with improved trial 
wave functions provides some assurance that FD-AFQMC can obtain 
accurate properties even in cases with more challenging 
electronic correlation -- in this case, a subtle competition between ionic and covalent electronic configurations as the single bond is broken.\cite{Mayhall2014}

However, we did encounter some unexpected challenges in converging the FD-AFQMC dipole moments along the FH dissociation coordinate.  For example, we observed greater sensitivity 
of FD-AFQMC to noise and potential loss of correlation at longer stretches, relative to our studies of equilibrium geometries.
With UHF trials, these challenges were overcome by employing smaller propagation times of only up to 20 Ha$^{-1}$, which was still sufficient to converge the dipole moments before any significant growth of stochastic noise due to the gradual loss of correlation between the primary and secondary systems.  In addition, we removed any outlier spikes in the energy difference trajectories of more than 10$^{-4}$ Ha.
Some FD-AFQMC calculations with CASSCF trial wavefunctions were found to be more sensitive to noise than those with UHF trials; we addressed this by using a slightly larger perturbing field strength of 10$^{-3}$ a.u. to successfully resolve  energy 
differences and obtain accurate dipole moments.
At very long bond lengths beyond 3.5 \r{A}, we encounter severe noise 
with CASSCF trials and incorrectly obtain non-zero dipole moments. 
Challenges with noise at long bond lengths have also been observed in the 
computation of forces\cite{Goings2025} and future work will 
seek to improve correlated sampling's performance in this regime.

In conclusion, we have introduced a finite-difference based approach within AFQMC 
to response properties and demonstrated that FD-AFQMC, with the 
aid of correlated sampling, can obtain highly accurate electric 
dipole moments even for relatively large molecular sizes and in regimes of strong electron correlation.
While simple Hartree-Fock trial wave functions are already sufficient 
in most situations, we have also shown that straightforward improvements in 
the trial quality can ensure accurate dipole moments in challenging cases such as the azulene molecule and the dissociation of hydrogen fluoride.
FD-AFQMC is scalable to system sizes  
beyond the reach of CCSD, CCSD(T), and AFQMC approaches to properties based on automatic differentiation, and we can routinely compute dipole moments for 
asymmetric molecules and large basis sets.
In future studies, our correlated sampling methodology could potentially be improved through a dynamic reweighting scheme\cite{Chen2023_cs} or 
the averaging of energy differences from multiple short trajectories,\cite{Shee2017} in combination with other finite difference formulations.
FD-AFQMC could also be extended to provide high quality reference values for dipole moments of electronically excited states.\cite{Damour2023}
We also plan to apply the correlated sampling FD-AFQMC protocol to other first-order response properties such as magnetic dipole moments, using gauge invariant atomic orbitals,\cite{Helgaker1991} and to other properties such as quadrupole moments and polarizabilities.

\section{Acknowledgments}
We thank Brad Ganoe for helpful discussions. 
This work was supported in part by the Big-Data Private-Cloud Research Cyberinfrastructure MRI-award funded by NSF under grant CNS-1338099 and by Rice University's Center for Research Computing (CRC). This research also used resources of the Oak Ridge Leadership Computing Facility, which is a DOE Office of Science User Facility supported under Contract DE-AC05-00OR22725. 
J. Shee acknowledges support from the Robert A. Welch
Foundation, Award Number C-2212.
This work used Expanse at the San Diego Supercomputing Center\cite{strande2021expanse} 
through allocation CHE240177 from the Advanced Cyberinfrastructure Coordination Ecosystem: Services and Support (ACCESS) program,\cite{boerner2023access}
which is supported by U.S. National Science Foundation grants 2138259, 2138286, 2138307, 2137603, and 2138296. 

\section{Supplementary information}
Cartesian coordinates of optimized structures, table of DFT dipole moment values

\bibliography{main}

\end{document}